\documentclass[aps,prl,showpacs,twocolumn,superscriptaddress]{revtex4}
\usepackage{epsfig}
\begin{document}


\title{Signature of clustering in quantum many body systems probed by the giant dipole resonance}

\author{Deepak Pandit}
\email[e-mail:]{deepak.pandit@vecc.gov.in}
\affiliation{Variable Energy Cyclotron Centre, 1/AF-Bidhannagar, Kolkata-700064, India}

\author{Debasish Mondal}
\affiliation{Variable Energy Cyclotron Centre, 1/AF-Bidhannagar, Kolkata-700064, India}
\affiliation{Homi Bhabha National Institute, Training School Complex, Anushaktinagar, Mumbai 400 094,India}

\author{Balaram Dey}
\email[Present address:Tata Institute of Fundamental Research, Mumbai 400 005, ]{India}
\affiliation{Variable Energy Cyclotron Centre, 1/AF-Bidhannagar, Kolkata-700064, India}

\author{Srijit Bhattacharya}
\affiliation{Department of Physics, Barasat Govt. College, Barasat, N 24 Pgs, Kolkata - 700124, India }

\author{S. Mukhopadhyay}
\affiliation{Variable Energy Cyclotron Centre, 1/AF-Bidhannagar, Kolkata-700064, India}
\affiliation{Homi Bhabha National Institute, Training School Complex, Anushaktinagar, Mumbai 400 094,India}

\author{Surajit Pal}
\affiliation{Variable Energy Cyclotron Centre, 1/AF-Bidhannagar, Kolkata-700064, India}

\author{A. De}
\affiliation{Department of Physics, Raniganj Girls' College, Raniganj-713358, India}

\author{S. R. Banerjee}
\email[e-mail:]{srb@vecc.gov.in}
\affiliation{Variable Energy Cyclotron Centre, 1/AF-Bidhannagar, Kolkata-700064, India}
\affiliation{Homi Bhabha National Institute, Training School Complex, Anushaktinagar, Mumbai 400 094,India}


\date{\today}

\begin{abstract}
The present experimental study illustrates how large deformations attained by nuclei due to cluster formation are perceived through the giant dipole resonance (GDR) strength function. The high energy GDR $\gamma$-rays have been measured from $^{32}$S at different angular momenta ($J$) but similar temperatures in the reactions $^{4}$He(E$_{lab}$=45MeV) + $^{28}$Si and $^{20}$Ne(E$_{lab}$=145MeV) + $^{12}$C. The experimental data at lower J ($\sim$ 10$\hbar$) suggests a normal deformation, similar to the ground state value, showing no potential signature of clustering. However, it is found that the GDR lineshape is fragmented into two prominent peaks at high J ($\sim$ 20$\hbar$) providing a direct measurement of the large deformation developed in the nucleus. The observed lineshape is also completely different from the ones seen for Jacobi shape transition at high $J$ pointing towards the formation of cluster structure in super-deformed states of $^{32}$S at such high spin. Thus, the GDR can be regarded as a unique tool to study cluster formation at high excitation energies and angular momenta.                

\end{abstract}
\pacs{24.30.Cz,24.60.Dr,25.70.Gh}
\maketitle

\section{INTRODUCTION}
The nucleus is a dynamic finite size system consisting of protons and neutrons, where their velocities can reach a significant fraction of the speed of light. The description of nuclear dynamics at such velocities is 
predominantly based on the concept of independent nucleons moving in a mean field potential.   
However, in spite of their independent random motions, the nucleons also have a propensity to congregate i.e. these nucleons possess correlations \cite{oer06, fre07}. 
The fact that the clustering of nucleons leads to the occurrence of molecular states in the atomic nuclei was already  realized in the earliest days of  nuclear physics study \cite{whee37}.
The nuclear cluster phase is considered as the transitional state between the crystalline and quantum liquid phases of a fermionic system, which is linked to the studies of the `nuclear pasta phase' in the crust of neutron stars \cite{ear12}.

The nuclear structure data in the s-d shell region provide a wonderful opportunity to study the clustering phenomena since the densities of the deformed one-body states often exemplify significant cluster structure in this region \cite{hor10}. Kimura and Horiuchi showed \cite{kim04} that the super deformed (SD) band members of $^{32}$S have a considerable amount of the $^{16}$O + $^{16}$O cluster component. The reaction calculations, using a deep $^{16}$O + $^{16}$O  potential appropriate to the entrance channel, suggested the existence of $^{16}$O + $^{16}$O cluster bands in $^{32}$S \cite{ohk02}. Similar SD band was obtained using the alpha-alpha double folding potential \cite{koc10}. Recently, evidence of such cluster formations was also predicted by the macroscopic-microscopic potential energy surface calculations for $^{32}$S \cite{ich11}. Ichikawa et al, emphasized  the inclusion of the rotational energy contribution and showed that the nuclear densities in the SD band become cluster-like at high angular momentum ($J$).
Experimentally, the inelastic scattering and the damped fragment yields, in the reaction $^{20}$Ne + $^{12}$C,  indicated  the survival of an orbiting dinuclear system \cite{sap79, san99, cha05}. 
It is now well known that these cluster structures are associated with strongly deformed shapes of nuclei.
The deformations, estimated from the respective $\alpha$-particle evaporation spectra in the reaction $^{20}$Ne + $^{12}$C, have been found to be much larger compared to normal deformation attained by hot rotating composites at similar excitation 
energies \cite{dey07a}.   
However, there has been no direct measurement of this deformation at high excitation energies and angular momenta. 

One of the probes to study this deformation experimentally at high excitation energies and angular momenta is the $\gamma$-decay from the giant dipole resonance (GDR) built on excited states. It is the prime example of collective nuclear vibration, which can be understood macroscopically as the out of phase oscillation between the protons and neutrons, and microscopically in terms of coherent particle-hole  excitations \cite{hara01, gaar92}. 
The GDR emission occurs early in the decay of excited nuclei and also couples directly 
with the nuclear shape degrees of freedom. 
Therefore, it is highly important to investigate experimentally the shapes of $^{32}$S at different angular momenta to study how cluster formations are manifested in the GDR strength function. 
The GDR lineshape should reveal direct information about the geometrical configurations of the nuclei, which can provide vital clues about the underlying mechanism to understand the nuclear structure and collective dynamics at extreme conditions of $T$ and $J$.   

It is very interesting to note that, in the long-wavelength limit, the E1 decay of the GDR $\gamma$-rays (isovector in nature) from self-conjugate nuclei is hindered since decays from same isospin ($I$) states are forbidden \cite{har86}.
The yield, however, increases in the presence of isospin mixing due to the weak Coulomb interaction characterized by isospin violating spreading width $\Gamma^\downarrow$ \cite{hara86, beh93}.
In this paper, we report on the measurement of GDR strength function for $^{32}$S at low and high $J$ in the reactions $^{4}$He + $^{28}$Si and $^{20}$Ne+ $^{12}$C, respectively and compare them to those obtained from the thermal shape fluctuation model. The Coulomb spreading width has also been estimated by populating nearby nucleus $^{31}$P in the reaction $^{4}$He + $^{27}$Al. We show that the GDR lineshape at low $J$ indicates normal deformation, whereas at higher J point towards large deformation due cluster formation.

\begin{figure}
\begin{center}
\includegraphics[height=9.0 cm, width=8.0 cm]{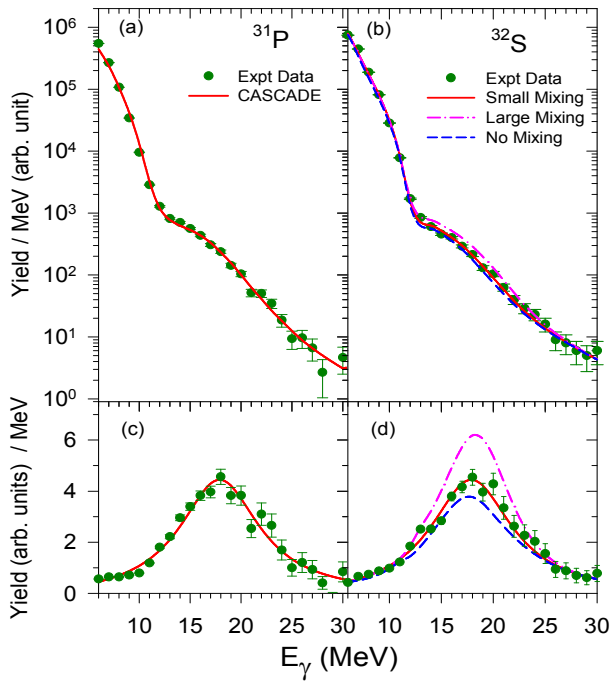}
\caption{\label{fig1} (color online) The experimental $\gamma$-spectra (symbols)
for $^{31}$P (a) and $^{32}$S (b) are shown for 46 MeV excitation energy along with the statistical model calculation
CASCADE plus bremsstrahlung. The corresponding linearized GDR strength functions are shown in  panels (c) and (d).}
\end{center}
\end{figure}

\section{EXPERIMENTAL DETAILS AND ANALYSIS}
The experiments were performed using the K-130 cyclotron at the Variable Energy Cyclotron Centre (VECC), Kolkata. The excited $^{32}$S nucleus was populated at lower $J$ in the reaction $^{4}$He(E$_{lab}$=45MeV) + $^{28}$Si. The initial excitation energy was 46.3 MeV while the critical angular momentum for fusion was 10$\hbar$. 
To extract the Coulomb spreading width, the $^{31}$P nucleus was also populated at the same excitation energy in the reaction $^{4}$He(E$_{lab}$=42MeV) + $^{27}$Al but with $I$ = 1/2 entrance channel. The high energy GDR $\gamma$-rays were detected using a part of the LAMBDA spectrometer \cite{supm07} arranged in a 7$\times$7 matrix. The spectrometer was kept at a distance of 50 cm from the target position at an angle of 90$^\circ$ to the beam direction. The GDR spectra were also measured at 55$^\circ$ and 125$^\circ$ to extract the bremsstrahlung slope parameter. Low energy $\gamma$-ray multiplicities were measured using the gamma multiplicity filter \cite{dipu3}. The 50-element filter was split into two blocks of 25 detectors each and was placed above and below the scattering chamber at a distance of 5 cm from the target center. The high energy $\gamma$-rays were separated from the neutron induced events employing the time of flight technique while the pile-up events were rejected using a pulse shape discrimination (PSD) technique by measuring the charge deposition over two integrating time intervals (50 ns and 2 $\mu$s) in each of the detectors. Finally, the high-energy spectra for higher folds of the multiplicity filter were generated in offline analysis following the cluster summing technique \cite{supm07, sri08}. The $^{32}$S nucleus was populated at higher $J$ in the reaction $^{20}$Ne(E$_{lab}$=145MeV) + $^{12}$C. The initial excitation energy was 73 MeV while the critical angular momentum for fusion was 21$\hbar$. 
The complete detector setup and experimental details can be found in Ref \cite{dipu1}.

The statistical model calculation was performed using a modified version of CASCADE in which isospin quantum number had been taken into account. The $\Gamma^\downarrow$ and the GDR parameters at low $J$ for $^{32}$S and $^{31}$P were extracted using $\chi^2$ method in the range of  12-24 MeV, following a recursive procedure described in detail in refs \cite{beh93, kin04, cor11}. The bremsstrahlung slope parameter was estimated by simultaneous fitting of the high energy $\gamma$-ray spectra and a$_1$(E$_\gamma$) coefficient, considering isotropic emission in a source frame moving with 0.6$v_{beam}$ \cite{kel99}. 
The extracted slope parameter was consistent with the bremstrahlung systematics \cite{nif90}.  
The set of best fit GDR parameters was found to be E$_{GDR}$ = 18.5 $\pm$ 0.2 MeV, $\Gamma_{GDR}$=9.5 $\pm$ 0.5 MeV and S$_{GDR}$ =1.1 $\pm$ 0.03. The extracted Coulomb spreading width for $^{32}$S was $\Gamma^\downarrow$=18 $\pm$ 12 keV and found to be consistent with the measurements carried out by other groups \cite{kin04}. The high energy spectra for $^{32}$S and $^{31}$P, along with the statistical model calculations plus bremsstrahlung, are shown in Fig \ref{fig1}. To emphasize the GDR region, the linearized GDR plots are also shown in the figure using the quantity 
F(E$_\gamma$)Y$^\textrm{exp}$(E$_\gamma$)/Y$^\textrm{cas}$(E$_\gamma$), where Y$^\textrm{exp}$(E$_\gamma$) and Y$^\textrm{cas}$(E$_\gamma$)  are the experimental and the best fit CASCADE spectra,
respectively, corresponding to Lorentzian function F(E$_\gamma$). The statistical calculations for $^{32}$S with zero mixing ($\Gamma^\downarrow$=0 keV) and large mixing ($\Gamma^\downarrow$=100 keV) are also compared in Fig \ref{fig1}. 
The average deformation was extracted from the GDR width using the emperical relation \cite{dipu4} and
ground state width as 7.5 MeV \cite{atlas}. The estimated deformation at low J ($\sim$ 10$\hbar$) is
$\beta$ =0.36, slightly higher than the ground state value ($\beta$ =0.31).  

\begin{figure}
\begin{center}
\includegraphics[height=7.0 cm, width=8.5 cm]{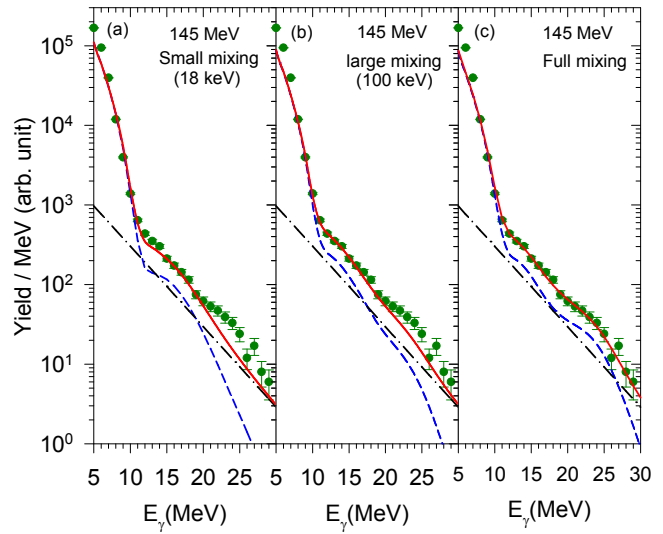}
\caption{\label{fig2} (color online) The statistcal model calculation plus bremsstrahlung (solid lines) are compared to experimental data (symbols) at 145 MeV incident energy using (a) small mixing (b) full mixing and  (c) including the pre-equlibrium effect. The individual CASCADE (dashed line) and bremsstrahlung (dashed-dotted line)
are also shown.}  
\end{center}
\end{figure}

It needs to be mentioned here that the data at higher angular momentum were analyzed earlier \cite{dipu1}. In this work, we  reanalzsed it using the isospin included CASCADE code.  
It has been seen experimentally and justified theoretically that $\Gamma^\downarrow$ remains constant with excitation energy \cite{har86}. It is also well known that the GDR width increases with excitation energy due to thermal fluctuations and angular momentum induced deformation but the E$_{GDR}$ remains constant \cite{hara01, dipu4, dipu2}.
Hence, the $^{32}$S  data, at higher J,  in the reaction $^{20}$Ne + $^{12}$C were tried to fit by varying only the GDR width.
Since the a$_1$ coefficient was not measured earlier for this reaction, the bremsstrahlung slope was estimated from the bremsstrahlung systematic \cite{nif90}. It can be seen that the data cannot be explained using a single Lorentzian in the statistical model calculation (Fig \ref{fig2}a). A second component in the higher energy region is evident ($\sim$25 MeV) even in the high-energy $\gamma$-ray spectrum. Therefore, the data were analysed considering two Lorentzian functions for the GDR in the CASCADE. 
Although one could fit the lower energy component with small isospin mixing, it was not possible to fit the higher energy component with $\Gamma^\downarrow$=18 keV.  Even for $\Gamma^\downarrow$=100 keV (which corresponds to large mixing), a strength function of 150 $\%$ of TRK sum rule also could not fit the higher energy component (Fig \ref{fig2}b).    
As a result, the data were analysed considering full mixing to extract the GDR components 
and is shown in Fig \ref{fig2}c. 
The extracted GDR parameters are E$_{GDR1}$ =14.7 $\pm$ 0.3 MeV, $\Gamma_{GDR1}$ =6.0 $\pm$ 0.8 MeV, S$_{GDR1}$ =0.33 $\pm$ 0.05, E$_{GDR2}$= 25.6 $\pm$ 0.8 MeV, $\Gamma_{GDR2}$=7.3 $\pm$ 1.3 MeV, S$_{GDR2}$=0.77 $\pm$ 0.09. 
The linearised GDR spectrum for E$_{lab}$ = 145 MeV is shown in Fig \ref{fig4}b using the quantity F(E$_\gamma$)Y$^\textrm{exp}$(E$_\gamma$)/Y$^\textrm{cas}$(E$_\gamma$).
The estimated deformation from the two GDR peaks is $\beta$ = 0.68 which corresponds to an axis ratio of 1:1.9.
In principle, the isospin mixing should be small for fusion-evaporation reaction. This is corroborated by the fact that small mixing ($\Gamma^\downarrow$ = 18 keV) can predict the experimentally obtained lower energy component (14.7 MeV) of the GDR spectra. In prolate deformed $^{32}$S nucleus, the observation of the low energy GDR component suggests that one should also have another broader component in the higher energy region (22-25 MeV) and isospin mixing should also be small for it. However, it can not be explained with small mixing which indicates that, apart from the $^{32}$S nucleus, the high energy component also has a contribution from much lighter mass nuclei. It may be noted that the extracted centroid and width of the second component are very similar to the $^{16}$O ground state values \cite{atlas}. Thus, the inability to explain the higher energy GDR component with standard parameters point towards the formation of cluster-like structures in a deformed $^{32}$S nucleus.

\section{RESULTS AND DISCUSSIONS}
One can conjecture that the origin of the high energy component ($\sim$ 25 MeV) is due to the emission of high-energy $\gamma$-rays from much lighter compound nuclei formed by incomplete fusion. However, in our earlier work on evaporation-residue-gated Jacobi shape transition \cite{drc12}, 
it was observed that the non-fusion events were accompanied by $\gamma$-rays in the range of 5-10 MeV
and were associated with low angular momentum events only. Since our measurement was biased towards the higher multiplicity events, it can be inferred that the $\gamma$-rays are emitted from a fully energy equilibrated composite.
Our earlier charge particle experiments \cite{cha05, dey07}, for the same reaction, clearly revealed that the damped fragments (Z=3-7) are emitted from a fully energy equilibrated composites and follow a 1/sin$\theta_{cm}$ angular dependence. 
Therefore, the contribution to the high energy component from incomplete fusion and deep-inelastic process can be completely ruled out. The charge particle studies  also revealed that this reaction proceeds via the long-lived di-nuclear orbiting mechanism at high angular momenta. For an orbiting mechanism, the system becomes trapped in a more deformed configuration than that of the compound nucleus and is inhibited from spreading into the compound nucleus states \cite{san99}. As it appears, the large yield of the higher GDR component is arises due to the nuclear orbiting process which leads to the cluster formation at higher $J$.

In general, the light nuclei are expected to undergo Jacobi shape transition, an abrupt change of shape from a non-collective oblate to a collectively rotating prolate or triaxial shape, at high angular momentum (J $\sim$ 17$\hbar$ for $^{32}$S). Experimentally, it is observed as a sharp low energy component ($\sim$ 10 MeV) in the GDR spectrum \cite{dipu1,maj04,drc12}. This peak arises due to the Coriolis splitting of the GDR frequency corresponding to the largest axis of a collectively rotating prolate when the frequencies are transformed from internal rotating coordinate frame to the laboratory frame \cite{gal85}. 
Interestingly, the Jacobi shape transition is also characterized by large deformation ($\beta$ $\sim$ 0.7).
However, no low energy component is observed at higher $J$ indicating that the Jacobi transition is not proceeding in this reaction. 
The possible reason can be due to the formation of the $^{16}$O + $^{16}$O cluster in $^{32}$S at high $J$ \cite{ich11} via the
nuclear orbiting mechanism due to the entrance channel. 
For such systems, the moment of inertia can be considered of a two-body freely rotating about an axis perpendicular to the symmetry axis rather than being a one-body rigid rotor. The moment of inertia of these molecular states has been found to be $\sim$ 1.5 times larger compared to the super deformed states \cite{sci09}. As a result,    
the angular frequency in this case would be much smaller reducing the effect of Coriolis splitting for a given $J$. 
In fact, considering the moment of inertia as a two-body freely rotating one, 
the calculated Coulomb barrier heights can explain the existing experimental data of molecular resonance states in the $^{16}$O + $^{16}$O reaction channel \cite{ich11}.

\begin{figure}
\begin{center}
\includegraphics[height=9.0 cm, width=8.3 cm]{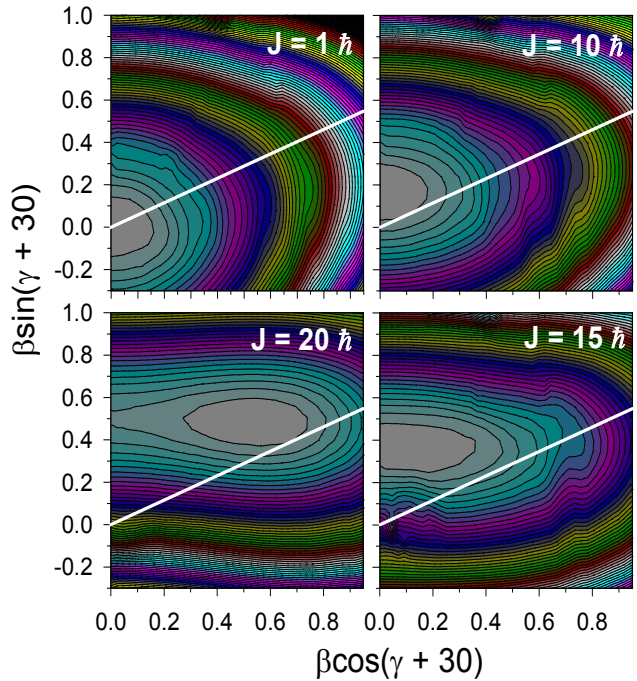}
\caption{\label{fig3} (color online) The free energy surfaces for $^{32}$S at $T$ = 2.0 MeV and different angular momenta.
The thick white solid line corresponds to $\gamma$=0.}
\end{center}
\end{figure}

\begin{figure}
\begin{center}
\includegraphics[height=7.5 cm, width=8.3 cm]{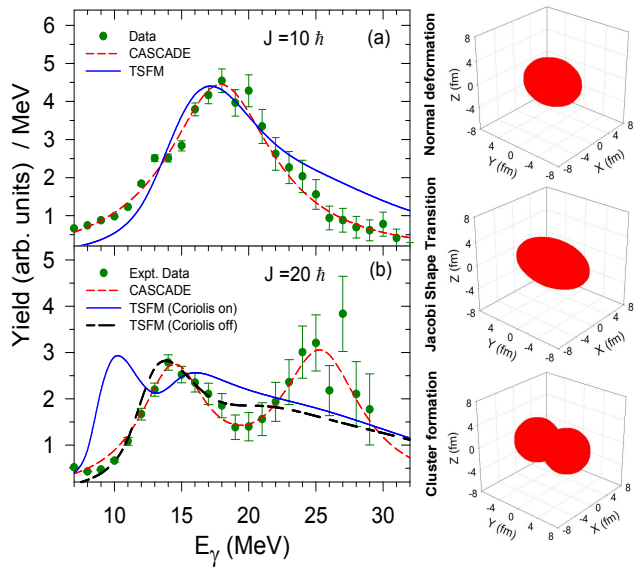}
\caption{\label{fig4} (color online) The experimental GDR strength functions (symbols) for $^{32}$S
are compared with TSFM calculation at low(a) and high(b) $J$. The normal (low $J$), Jacobi transition and cluster formation shapes are also shown on the right side.}
\end{center}
\end{figure}

Finally, the theoretical GDR lineshapes for the $^{32}$S nucleus were also generated based on the thermal shape fluctuation model (TSFM) at both low and high angular momenta \cite{alh90,aru04, dub05}. The average temperature of the nucleus associated with the GDR decay was estimated from $\left\langle T \right\rangle$=[($\overline{E^*}$ - $\overline{E}_{rot}$ - E$_{GDR}$)/$a(\overline{E^*})$]$^{1/2}$  using the CASCADE code. $\overline{E^*}$ is the average of the excitation energy weighed over the daughter nuclei for the $\gamma$ emission in the GDR region and $\overline{E}_{rot}$ is the average rotational energy. The  level density parameter was taken as A/8. The temperature corresponding to 45 MeV incident energy was 
2.0 $\pm$ 0.2 MeV while for 145 MeV incident energy it was 2.3 $\pm$ 0.4 MeV. 
The free energy surfaces for the TSFM calculation were estimated using the relation 
F(T,J;$\beta$,$\gamma$) = F(T,J=0;$\beta$,$\gamma$) + $\frac{J(J+1)\hbar^2}{2(\omega.I.\omega)}$
where $\omega.I.\omega$ = $I_{xx}sin^2\theta cos^2\phi+ I_{yy}sin^2\theta sin^2\phi + I_{zz}cos^2\theta$
is the moment of inertia about the rotation axis $\omega$.
F(T,J=0;$\beta$,$\gamma$) is the non-rotating part and I$_{xx}$, I$_{yy}$, I$_{zz}$ are the principal rigid body moments of inertia. It was observed that, at these temperatures, the shell corrections (included in the calculation) were small, and F was predominantly given by the properties of a rotating liquid drop \cite{dipu1}. 
The free energy surfaces at T = 2 MeV for different $J$ are shown in Fig \ref{fig3}. 
The TSFM calculations at corresponding $J$ and $T$ are compared with the experimental data in Fig \ref{fig4}.  
As can be seen, the TSFM calculation predicted a slightly larger width, as expected \cite{dipu2}, but 
represented the overall nature of the experimental lineshape at low $J$.
On the other hand, the data are in complete disagreement with the TSFM calculation at higher $J$. The
calculation shows a sharp 10 MeV peak characteristics of the Jacobi shape transition but no such component is observed
in the experiment. The GDR lineshapes were also generated at different angular momenta but it failed completely to explain the experimental data. We should mention here that the TSFM calculation does not take into account the cluster formation (higher order deformation) in the equilibrated nuclei. However, a calculation was performed by switching off the Coriolis splitting of the GDR components. Interestingly, the theoretical lineshape quite well explains the low energy component highlighting that the Coriolis effect is indeed small at high J, as expected for cluster formation. But the calculation fails to represent the high energy part of the spectrum as it does not take into account the GDR component due to clusterization which is beyond the scope of the present work.  
As it appears, the possible reason for not observing the Jacobi transition at high $J$ primarily seems  
to be due to the formation of the $^{16}$O + $^{16}$O component in $^{32}$S via the nuclear orbiting process
due to the target-projectile combination as predicted by the authors of Ref\cite{ich11}. 
The Jacobi transition, cluster formation and normal deformation (at low J)  shapes are shown in Fig \ref{fig4}. 
Very recently, the GDR decay from $^{31}$P in the reaction $^{19}$F + $^{12}$C has been measured at high J where an enhanced
yield at around 10 MeV has been observed suggesting the onset of the Jacobi transition in the nearby nucleus of $^{32}$S \cite{bala16}.
Thus, all the experimental results, though indirect, clearly suggest that the $^{32}$S nucleus is not undergoing normal fusion evaporation mechanism and point towards the formation of cluster structure at high J rather convincingly.
In the future, it will be an intriguing study to measure the GDR $\gamma$-decay from nearby self-conjugate nuclei 
$^{28}$Si and $^{36}$Ar, and search for the absence of the Jacobi transition which will establish 
GDR as a probe to study clustering in atomic nuclei at high $T$ and $J$.
From the theoretical point of view, it will be an interesting study to generate the GDR lineshapes due to cluster structures at high excitation as was performed recently \cite{he14} for $^{12}$C and $^{16}$O at respective alpha decay thresholds.

\section{SUMMARY}
In summary, the GDR $\gamma$-rays from $^{32}$S were experimentally measured both at low and high $J$ to study the 
manifestation of clusterization via the GDR spectra.
Another experiment was performed to extract the Coulomb spreading width by populating $^{31}$P at the same excitation energy to estimate the isospin mixing. At low $J$, the GDR lineshape suggests a normal deformation pointing towards the usual compound nucleus evolution.  
On the other hand, the nucleus is not proceeding via the usual fusion evaporation process at high J. The GDR lineshape suggests  superdeformation ($\beta$ $\sim$ 0.7), completely different from the Jacobi transition lineshape, which largely points toward cluster formation in super deformed states of $^{32}$S.    
 
\section{ACKNOWLEDGMENTS}
The authors gratefully acknowledge helpful discussions with  M. N. Harakeh. The authors would like to thank  A. Corsi for providing the isospin included CASCADE code originally obtained from  M. Kicinska-Habior.


\begin{thebibliography}{99}



\bibitem{oer06}  W. von Oertzen, M. Freer, Y. K. -En'yo, Physics Reports {\bf 432}, 43 (2006).
\bibitem{fre07}  M. Freer, Rep. Prog. Phys. {\bf 70}, 2149 (2007).
\bibitem{whee37} John A. Wheeler, Phys. Rev. {\bf 52}, 1107 (1937). 
\bibitem{ear12}  J. P. Ebran, E. Khan, T. Niksic and D Vretenar, Nature {\bf 487}, 341 (2012).
\bibitem{hor10}  H. Horiuchi, Clusters in Nuclei - Vol.1, ed, C Beck, Lecture Notes in Physics {\bf 818}, 57 (2010).
\bibitem{kim04}  M. Kimura and H. Horiuchi, Phys. Rev. C {\bf 69}, 051304(R) (2004).
\bibitem{ohk02}  S. Ohkubo and K. Yamashita, Phys. Rev. C {\bf 66}, 021301(R) (2002).
\bibitem{koc10}  G. Kocak, M. Karakoc, I. Boztosun, and A. B. Balantekin, Phys. Rev. C {\bf 81}, 024615 (2010).
\bibitem{ich11}  T. Ichikawa, Y. K. -En'yo and P. Moller, Phys. Rev. C {\bf 83}, 054319 (2011).
\bibitem{sap79}  D. Shapira et. al., Phys. Rev. Lett. {\bf 43}, 1781 (1979). 
\bibitem{san99}  S.J. Sanders, A. Szanto de Toledo, C. Beck, Physics Reports {\bf311} (1999) 487.
\bibitem{cha05}  C. Bhattacharya et. al., Phys. Rev. C {\bf 72}, 021601 (2005).
\bibitem{dey07a}  A. Dey et. al., Phys. Rev. C {\bf 74}, 044605 (2006).
\bibitem{hara01} M. N. Harakeh and A. van der Woude, Giant Resonances, Fundamental High-frequency Modes of
                 Nuclear Excitation, Clarendon Press, Oxford, 2001.
\bibitem{gaar92} J. J. Gaardhoje, Ann. Rev. Nucl. Part. Sci. {\bf 42} 483 (1992).
\bibitem{har86}  H. L. Harney, A Richter and H. A. Weidenmuller, Rev. Mod. Phys. 58, (1986) 607.
\bibitem{hara86} M. N. Harakeh et. al., Phys. Lett. B {\bf 176}, 297 (1986).
\bibitem{beh93}  J. A. Behr et. al., Phys. Rev. Lett. {\bf 70}, 3201 (1993).
\bibitem{supm07} S. Mukhopadhayay et al, Nucl. Instr. and Meth. A {\bf 582}, 603 (2007).
\bibitem{dipu3}  Deepak Pandit et al, Nucl. Instr. and Meth. A {\bf 624}, 148 (2010).
\bibitem{sri08}  Srijit Bhattacharya et al., Phys. Rev. C {\bf 77},  024318 (2008).
\bibitem{dipu1}  Deepak Pandit et. al., Phys. Rev. C {\bf 81}, 061302(R) (2010).
\bibitem{kin04}  M. Kicinska-Habior et al, Nucl. Phys. A {\bf 731}, 138 (2004).
\bibitem{cor11}  A. Corsi et. al., Phys. Rev. C {\bf 84}, 041304(R) (2011).
\bibitem{kel99}  M. P. Kelly et. al., Phys. Rev. Lett. {\bf 82}, 3404 (1999).
\bibitem{nif90}  H. Nifennecker, J.A. Pinston, Annu. Rev. Nucl. Part. Sci. {\bf 87}, 113 (1990).
\bibitem{dipu4}  Deepak Pandit et. al., Phys. Rev. C {\bf 87}, 044325 (2013).
\bibitem{atlas}  A. V. Varlamov, V. V. Varlamov, D. S. Rudenko, and M. E. Stepanov, Atlas of Giant Dipole Resonances,                            INDC(NDS)-394, 1999 (unpublished); JANIS database.
\bibitem{dipu2}  Deepak Pandit et. al., Phys. Lett. B {\bf 713}, 434 (2012).
\bibitem{drc12}  D.R. Chakrabarty et. al., Phys. Rev. C {\bf 85}, 044619 (2012).
\bibitem{dey07}  A. Dey et. al., Phys. Rev. C {\bf 76}, 034608 (2007).
\bibitem{maj04}  A. Maj et. al., Nucl. Phys. {\bf A731}, 319 (2004).
\bibitem{gal85}  M. Gallardo et. al., Nucl. Phys. A {\bf 443}, 415 (1985).
\bibitem{sci09}  Wagner Sciani et. al., Phys. Rev. C {\bf 80}, 034319 (2009). 
\bibitem{alh90}  Y. Alhassid and B. Bush, Phys. Rev. Lett. 65, 2527 (1990).
\bibitem{aru04}  P. Arumugam, G. Shanmugam and S. K. Patra, Phys. Rev. C {\bf 69}, 054313 (2004).
\bibitem{dub05}  N. Dubray, J. Dudek and A. Maj, Acta Physica Polonica B Vol {\bf 36}, 1161 (2005). 
\bibitem{bala16} Balaram Dey et. al., DAE-BRNS Symposiun on Nuclear Physics Vol {\bf 61} (2016) 100.
\bibitem{he14}   W. B. He et. al., Phys. Rev. Lett. {\bf 113}, 032506 (2014).



\end{thebibliography}
\end{document}